\documentclass{article}
\pdfoutput=1
\usepackage{spconf}
\usepackage{amsmath}
\usepackage{graphicx}
\usepackage{xcolor}
\usepackage{xspace}
\usepackage{amsfonts}
\usepackage{url} 
\usepackage[colorlinks=true]{hyperref} 
\usepackage[ruled]{algorithm2e}
\definecolor{NewTextBG}{rgb}{0.8,0.8,1.00}

\newcommand{\best}[1]{\textbf{\color{blue}#1}\xspace}
\newcommand{\BlackBox}{\rule{1.5ex}{1.5ex}}  

\newcommand{\opt}[1]{\ensuremath{\hat{#1}}}

\def\reals{\ensuremath{\mathbb{R}}}

\renewcommand{\vec}[1]{\ensuremath{\mathbf{\MakeLowercase{#1}}}}
\newcommand{\mat}[1]{\ensuremath{\mathbf{\MakeUppercase{#1}}}}

\newcommand{\st}{\ensuremath{\quad\mathrm{s.t.}\quad}}
\newcommand{\norm}[1]{\ensuremath{\left\|#1\right\|}}
\newcommand{\cost}[1]{\ensuremath{\ell_{#1}}\xspace}
\newcommand{\abs}[1]{\ensuremath{\left|#1\right|}}

\newcommand{\refeq}[1]{(\ref{#1})}

\def\Gaussian{\ensuremath{\mathcal{N}}}
\def\Exponential{\ensuremath{\mathrm{Exp}}}
\def\Bernoulli{\ensuremath{\mathrm{Ber}}}
\def\Laplacian{\ensuremath{\mathcal{L}}}
\def\LG{\ensuremath{\mathrm{LG}}}

\def\sgn{\mathrm{sgn}}

\def\dict{d}
\def\dictm{\mat{\dict}}
\def\dictv{\vec{\dict}}

\def\data{y}
\def\datam{\mat{\data}}
\def\datav{\vec{\data}}

\def\err{e}
\def\errm{\mat{\err}}
\def\errv{\vec{\err}}
\def\errrv{\epsilon}

\def\noiserv{\eta}

\def\coef{a}
\def\coefm{\mat{\coef}}
\def\coefv{\vec{\coef}}
\def\coefrv{\alpha}

\def\supp{z}

\def\suppv{\vec{\supp}}
\def\supprv{\zeta}

\def\sign{s}

\def\signv{\vec{\sign}}
\def\signrv{\phi}

\def\corr{g}

\def\corrv{\vec{\corr}}

\def\canon{\omega}

\def\corrv{\vec{\corr}}

\def\val{v}

\def\valv{\vec{\val}}
\def\valrv{\nu}

\def\aux{u}

\def\ndims{m} 
\def\natoms{p}
\def\nsamples{n}
\def\nclasses{c}

\def\di{i} 
\def\si{j} 
\def\ai{k} 

\def\supsize{\gamma}

\title{\MakeUppercase{Sparse coding and dictionary learning based on the MDL principle}}
\name{Ignacio Ram\'{i}rez and Guillermo Sapiro\thanks{Work supported by NSF, NGA, ARO, ONR, and NSSEFF.}}
\address{Department of Electrical and Computer Engineering, University of Minnesota}
\begin{document}
\ninept
\maketitle
\begin{abstract}
  The power of sparse signal coding with learned dictionaries has been
  demonstrated in a variety of applications and fields, from signal
  processing to statistical inference and machine learning. However, the
  statistical properties of these models, such as underfitting or
  overfitting \emph{given} sets of data, are still not well characterized in
  the literature. This work aims at filling this gap by means of the Minimum
  Description Length (MDL) principle -- a well established
  information-theoretic approach to statistical inference. The resulting
  framework derives a family of efficient sparse coding and modeling
  (dictionary learning) algorithms, which by virtue of the MDL principle,
  are completely parameter free. Furthermore, such framework allows to
  incorporate additional prior information in the model, such as Markovian
  dependencies, in a natural way.  We demonstrate the performance of the
  proposed framework with results for image denoising and classification
  tasks.
\end{abstract}

\section{Introduction}
\label{sec:introduction}
Sparse models are by now well established in a variety of fields and
applications, including signal processing, machine learning, and statistical
inference, e.g. \cite{candes06,aharon06,mairal08d} and references therein.

When sparsity is a modeling device and not a hypothesis about the nature of
the analyzed signals, parameters such as the desired sparsity in the
solutions, or the size of the dictionaries to be learned, play a critical
role in the effectiveness of sparse models for the tasks at hand. However,
lacking theoretical guidelines for such parameters, published applications
based on learned sparse models often rely on either cross-validation or
ad-hoc methods for learning such parameters (an exception for example
being the Bayesian approach, e.g.,~\cite{carin11})
. Clearly, such techniques can be impractical and/or ineffective in many
cases.

At the bottom of this problem lie fundamental questions such as: how rich or
complex is a sparse model? how does this depend on the required sparsity of
the solutions, or the size of the dictionaries? what is the best model for a
given data class?  A possible objective answer to such questions is provided
by the \emph{Minimum Description Length principle} (MDL)~\cite{rissanen84},
a general methodology for assessing the ability of statistical models to
capture regularity from data. The MDL principle is often regarded as a
practical implementation of the Occam's razor principle, which states that,
given two descriptions for a given phenomenon, the shorter one is usually
the best. In a nutshell, MDL equates ``ability to capture regularity'' with
``ability to compress'' the data, and the metric with which models are
measured in MDL is \emph{codelength} or \emph{compressibility}.

The idea of using MDL for sparse signal coding was explored in the
context of wavelet-based image denoising
\cite{saito94,moulin00}. These pioneering works were restricted to
denoising using fixed orthonormal basis (wavelets). In addition, the
underlying probabilistic models used to describe the transform coefficients,
which are the main technical choice to make when applying MDL to a problem,
were not well suited to the actual statistical properties of the modeled
data (image encoding coefficients), thus resulting in poor
performance. Furthermore, these works did not consider the critical effects
of quantization in the coding, which needs to be taken into
account when working in a true MDL framework (a useful model needs to be
able to compress, that is, it needs to produce actual codes which are
shorter than a trivial description of the data). Finally, these models are
designed for noisy data, with no provisions for the important case of
noiseless data modeling, which addresses the errors due to deviations from
the model, and is critical in many applications such as classification.

The framework presented in this work addresses all of the above issues in a
principled way: i) Efficient codes are used to describe the encoded data;
ii) Deviations from the model are taken into account when modeling
approximation errors, thus being more general and robust than traditional
sparse models; iii) Probability models for (sparse) transform coefficients
are corrected to take into account the high occurrence of zeros; iv)
Quantization is included in the model, and its effect is treated rigorously;
v) Dictionary learning is formulated in a way which is consistent with the
model's statistical assumptions.  At the theoretical level, this brings us a
step closer to the fundamental understanding of sparse models and brings a
different perspective, that of MDL, into the sparse world. From a practical
point of view, the resulting framework leads to coding and modeling
algorithms which are completely parameter free and computationally
efficient, and practically effective in a variety of applications.

Another important attribute of the proposed family of models is that prior
information can be easily and naturally introduced via the underlying
probability models defining the codelengths. The effect of such priors can
then be quickly assessed in terms of the new codelengths obtained. %
For example, Markovian dependencies between  sparse codes of adjacent image patches can be easily incorporated.

\section{Background on sparse models}
\label{sec:background}

%
Assume we are given $\nsamples$ $\ndims$-dimensional data samples ordered as
columns of a matrix $\datam=[\datav_1|\datav_2|\ldots|\datav_\nsamples] \in
\reals^{\ndims{\times}\nsamples}$. Consider a linear model for $\datam$,
$\datam = \dictm\coefm + \errm,$ where
$\dictm=[\dictv_1|\dictv_2|\ldots|\dictv_\natoms]$ is an
$\ndims{\times}\natoms$ dictionary consisting of $\natoms$ atoms,
$\coefm=[\coefv_1|\coefv_2|\ldots|\coefv_\nsamples] \in \reals^{\natoms{\times}\nsamples}$ is a matrix of
coefficients where each $\si$-th column $\coefv_\si$
specifies the linear combination of columns of $\dictm$ that approximates
$\datav_\si$, and $\errm=[\errv_1|\errv_2|\ldots|\errv_\nsamples] \in \reals^{\ndims{\times}\nsamples}$ is a matrix
of approximation errors.  We say that the model is \emph{sparse} if, for all
or most $\si=1,\ldots,\nsamples$, we can achieve $\norm{\errv_\si}_2 \approx
0$ while requiring that only a few coefficients $\supsize \ll \natoms$ in
$\coefv_\si$ can be nonzero. 

The problem of sparsely representing $\datam$ in terms of
$\dictm$, which we refer to as the \emph{sparse coding problem}, can be
written as
\begin{equation}
{\opt{\coefv}_\si} \!=\! \arg\min_{\coefv} \norm{\coefv}_0 
\!\st\! \norm{\datav_\si\!-\!\dictm\coefv}_2 \leq \varepsilon,\;
 \si=1,\ldots,\nsamples,%
\label{eq:sparse-coding}%
\end{equation}
where $\norm{\coefv}_0=\supsize$ is the pseudo-norm that counts the number
of nonzero elements in a vector $\coefv$, and $\varepsilon$ is some small
constant. There is a body of results showing that the problem
\refeq{eq:sparse-coding}, which is non-convex and NP-hard, can be solved
exactly when certain conditions on $\dictm$ and $\coefm$ are met, by either
well known greedy methods such as Matching Pursuit \cite{mallat93}, or by
solving a convex approximation to \refeq{eq:sparse-coding}, commonly
referred to as Basis Pursuit (BP) \cite{chen98},
\begin{equation}
\opt{\coefv}_\si = \arg\min_{\coefv} \norm{\coefv}_1 
\!\!\st\! \norm{\datav_\si-\dictm\coefv}_2 \leq \varepsilon,\;
 \si=1,\ldots,\nsamples.
\label{eq:l1-sparse-coding}
\end{equation}
When $\dictm$ is included as an optimization variable, we refer to the
resulting problem as \emph{sparse modeling}. This problem is often written
in unconstrained form,
\begin{equation}
  (\opt\dictm,\opt\coefm) = 
\arg\min_{\dictm,\coefm} \sum_{\si=1}^{\nsamples}
{\frac{1}{2}\norm{\datav_\si-\dictm\coefv_\si}_2^2 + \lambda \norm{\coefv_\si}_1},
\label{eq:l1-sparse-modeling}
\end{equation}
where $\lambda > 0$ is an arbitrary constant. The problem in this case is
non-convex in $(\dictm,\coefm)$, and one must be contempt with finding local
minima. Despite this drawback, in recent years, models learned by
(approximately) minimizing \refeq{eq:l1-sparse-modeling} have shown to be
very effective for signal analysis, leading to state-of-the-art results in
several applications such as image restoration and classification.

\subsection{Model complexity of sparse models}

In sparse modeling problems where $\dictm$ is learned, parameters such as
the desired sparsity $\supsize$, the penalty $\lambda$ in
\refeq{eq:l1-sparse-modeling}, or the number of atoms $\natoms$ in $\dictm$,
must be chosen individually for each application and type of data to produce
good results. In such cases, most sparse modeling techniques end-up using
cross-validation or ad-hoc techniques to select these critical
parameters. An alternative formal path is to postulate a Bayesian model
where these parameters are assigned prior distributions, and such priors are
adjusted through learning. This approach, followed for example in
\cite{carin11}, adds robustness to the modeling framework, but leaves
important issues unsolved, such as providing objective means to compare
different models (with different priors, for example). The use of
Bayesian sparse models implies having to repeatedly solve possibly costly
optimization problems, increasing the computational burden of the
applications.

In this work we propose to use the MDL principle to formally tackle the
problem of sparse model selection. The goal is twofold: for sparse coding
with fixed dictionary, we want MDL to tell us the set of coefficients that
gives us the shortest description of a given sample. For dictionary
learning, we want to obtain the dictionary which gives us the shortest
average description of all data samples (or a representative set of samples
from some class). A detailed description of such models, and the
coding and modeling algorithms derived from them, is the subject of the next
section.

\section{MDL-based sparse coding and modeling framework}
\label{sec:contribution}

Sparse models break the input data $\datam$ into three parts: a dictionary
$\dictm$, a set of coefficients $\coefm$, and a matrix of reconstruction
errors.  In order to apply MDL to a sparse model, one must provide
codelength assignments for these components, $L(\coefm)$, $L(\dictm)$ and
$L(\errm)$, so that the total codelength $L(\datam) =
L(\errm)+L(\coefm)+L(\dictm)$ can be computed. In designing such models, it
is fundamental to incorporate as much prior information as possible so that
no cost is paid in learning already known statistical features of the data,
such as invariance to certain transformations or symmetries. Another feature
to consider in sparse models is the predominance of zeroes in $\coefm$.  In
MDL, all this prior information is embodied in the probability models used
for encoding each component. What follows is a description of such models.

{\noindent\bf{Sequential coding}--} In the proposed framework, $\datam$ is
encoded sequentially, one column $\datav_\si$ at a time, for
$j=1,2,\ldots,\nsamples$, possibly using information (including
dependencies) from previously encoded columns. However, when encoding each
column $\datav_\si$, its sparse coefficients $\coefv_\si$ are modeled as and
IID sequence (of fixed length $\natoms$).

{\noindent\bf{Quantization}--} To achieve true compression, the finite
precision of the input data $\datam$ must be taken into account.  In the
case of digital images for example, elements from $\datam$ usually take only
$256$ possible values (from $0$ to $255$ in steps of size
$\delta_\data=1$). Since there is no need to encode $\errm$ with more
precision than $\datam$, we set the error quantization step to
$\delta_\err=\delta_\data$. As for $\dictm$ and $\coefm$, the corresponding
quantization steps $\delta_\coef$ and $\delta_\dict$ need to be fine enough
to produce fluctuations on $\dictm\coefm$ which are smaller than the
precision of $\datam$, but not more. Therefore, the actual distributions
used are discretized versions of the ones discussed below.

{\noindent\bf{Error}--} The elements of $\errm$ are encoded with an IID
model where the random value of a coefficient (represented by an
r.v. $\errrv$) is the linear superposition of two effects:
$\errrv=\hat{\errrv}+\noiserv$, where $\noiserv \sim
\Gaussian(0,\sigma^2_\noiserv)$, $\sigma^2_\noiserv$ assumed known, models
noise in $\datam$ due to measurement and/or quantization, and $\hat{\errrv}
\sim \Laplacian(0,\theta_\errrv)$ is a zero mean, heavy-tailed (Laplacian)
error due to the model. The resulting distribution for $\errrv$, which is
the convolution of the involved Laplacian and Gaussian distributions, was
developed in \cite{ramirez10dude} under the name ``LG.''  This model will be
referred to as $P_{\errrv}(\cdot;\sigma^2_\noiserv,\theta_\errrv)$
hereafter.

{\noindent\bf{Sparse code}--} Each coefficient in $\coefm$ is modeled as the
product of three (non-independent) random variables (see also \cite{carin11}
for a related model), $\coefrv = \supprv\signrv(\valrv+\delta_\coef)$, where
$\supprv \sim \Bernoulli(\rho)$ is a support indicator, that is, $\supprv=1$
implies $\coefrv \neq 0$, $\signrv =\sgn(\coefrv)$, and
$\valrv=\max\{\abs{\coefrv}-\delta_\coef,0\}$ is the absolute value
of $\coefrv$ corrected for the fact that $\valrv\geq \delta_\coef$ when
$\supprv=1$. Conditioned on $\supprv=0$, $\signrv=\valrv=0$ with probability
$1$. Conditioned on $\supprv=1$, we assume $\signrv \sim \Bernoulli(1/2)$,
and $\valrv$ to be $\Exponential(\theta_{\valrv})$. Note that, with these
choices, $P(\signrv\valrv|\supprv=1)$ is a Laplacian, which is a standard
model for transform (e.g., DCT, Wavelet) coefficients. The probability
models for the variables $\supprv$ and $\valrv$ will be denoted as
$P_{\supprv}(\cdot;\rho)$ and $P_{\valrv}(\cdot;\theta_{\valrv})$
respectively.

{\noindent\bf{Dictionary}--} We assume the elements of $\dictm$ to be
uniformly distributed on $[-1,1]$. Following the standard MDL recipe for
encoding model parameter values learned from $\nsamples$ samples, we use a
quantization step $\delta_\dict=\nsamples^{-1/2}$ \cite{rissanen84}. For
these choices we have $L(\dictm)=\ndims\natoms\log_2\nsamples$, which
does not depend on the element values of $\dictm$ but only on the number of
atoms $\natoms$ and the size of $\datam$. Other possible models which impose
structure in $\dictm$, such as smoothness in the atoms, are natural to the
proposed framework and will be treated in the extended version of this work.

\subsection{Universal models for unknown parameters}

The above probability models for the error $\errrv$, support $\supprv$ and
(shifted) coefficient magnitude $\valrv$ depend on parameters which are
not known in advance. In contrast with Bayesian approaches,
cross-validation, or other techniques often used in sparse modeling, modern
MDL solves this problem efficiently by means of the so called
\emph{universal probability models} \cite{rissanen84}. In a nutshell,
universal models provide optimal codelengths using probability distributions
of a known family, with unknown parameters, thus generalizing the classic
results from Shannon theory \cite{cover06}.

Following this, we substitute $P_{\supprv}(\cdot;\rho)$,
$P_{\errrv}(\cdot;\sigma^2_\noiserv,\theta_\errrv)$ and
$P_{\valrv}(\cdot;\theta_\val)$ with corresponding universal models.  For
describing a given support $\suppv$, we use an enumerative code
\cite{cover73}, which first describes the size of the support $\supsize$
with $\log_2 \natoms$ bits, and then the particular arrangement of non-zeros
in $\suppv$ using $\log_2 {\natoms \choose \supsize}$ bits.
$P_{\errrv}(\cdot;\sigma^2_\noiserv,\theta_\errrv)$ and
$P_{\valrv}(\cdot;\theta_\val)$ are substituted by corresponding universal
mixture models, one of the possibilities dictated by the theory of universal
modeling,
\begin{align*}
Q_{\errrv}(\aux;\sigma^2_\noiserv) &= \!\! \int_{0}^{+\infty}{\!\!\!\!\!\!\!\!w_\errrv(\theta)P_{\errrv}(\aux;\sigma^2_\noiserv,\theta)d\theta},\\
Q_\valrv(\aux) &=\!\!
  \int_{0}^{+\infty}{\!\!\!\!\!\!\!\!w_\valrv(\theta)P_{\valrv}(\aux;\theta)d\theta},
\end{align*}
where the mixing functions $w_\err(\theta)$ and $w_\coef(\theta)$  are Gamma distributions
(the conjugate prior for the exponential distribution),
$w(\theta|\kappa,\beta) =
{\Gamma(\kappa)}^{-1}\theta^{\kappa-1}\beta^{\kappa}e^{-\beta\theta},\;
\theta \in \reals^{+}$ with \emph{fixed} parameters
$(\kappa_\errrv,\beta_\errrv)$ and $(\kappa_\valrv,\beta_\valrv)$ respectively. The
resulting Mixture of
Exponentials (MOE) distribution $Q_\valrv(\cdot)$, is given by (see \cite{ramirez10tip} for details),
$
  Q_{\valrv}(\aux|\beta_\valrv,\kappa_\valrv) = \kappa_\valrv\beta_\valrv^{\kappa_{\valrv}}(\aux+\beta_\valrv)^{-(\kappa_\valrv+1)},\;u \in \reals^{+}.
$
Observing that the convolution and the convex mixture operations that result
in $Q_\errrv(\cdot;\sigma^2_\noiserv)$ are interchangeable (both integrals
are finite), it turns out that $Q_\errrv(\cdot;\sigma^2_\noiserv)$ is a
convolution of a MOE (of hyper-parameters $(\kappa_\errrv,\beta_\errrv)$)
and a Gaussian with parameter $\sigma^2_\noiserv$. Thus, although the
explicit formula for this distribution, which we call MOEG, is cumbersome,
we can easily combine the results in \cite{ramirez10dude} for the LG, and
for MOE in \cite{ramirez10tip} to perform tasks such as parameter estimation
within this model. Note that the universality of these mixture models does
not depend on the values of the hyper-parameters, and their choice has
little impact on their overall performance. Here, guided by
\cite{ramirez10tip}, we set $\kappa_\valrv=\kappa_\errrv=3.0$,
$\beta_\valrv=\beta_\errrv=\delta_\coef$.

Following standard practice in MDL, the \emph{ideal} Shannon code is used to
translate probabilities into codelengths. Under this scheme, a sample value
$\aux$ with probability $P(\aux)$ is assigned a code with length $L(\aux) =
-\log P(\aux)$ (this is an ideal code because it only specifies a
codelength, not a specific binary code, and because the codelengths produced
can have a fractional number of bits). Then, for a given error residual
$\errv$ and coefficients magnitude vector
$\valv=[\max\{\abs{\coef_\di}-\delta_\coef,0\}]_{\di=1,\ldots,\ndims}$, the
respective ideal codelengths will be $L_{\errrv}(\errv) =
\!\sum_{\di=1}^{\ndims}{\!\!-\log_2 Q_\errrv(\err_{\di})}$ and $
L_{\valrv}(\valv) = \!\sum_{{\ai=1}\\{\suppv_{\ai}=1}}^{\natoms}{\!\!-\log_2
  Q_\valrv(\val_{\ai})}.$ Finally, following the assumed $\Bernoulli(1/2)$
model, the sign of each non-zero element in $\coefv$ is encoded using 1 bit,
for a total of $\gamma$ bits.


\subsection{MDL-based sparse coding algorithms}
\label{sec:coding}

The goal of a coding algorithm in this framework is to obtain, for each
$\si$-th sample $\datav_\si$, a vector $\coefv_\si$ which minimizes its
description,
\begin{equation}
\opt{\coefv}_\si = \arg\min_{\coefv} L(\errv,\coefv) = L_{\errrv}(\errv_\si) + L_{\supprv}(\suppv) + L(\signv|\suppv) + L_{\valrv}(\valv|\suppv).
\label{eq:total-codelength}
\end{equation}
As it happens with most model selection algorithms, considering the support
size ($\supsize=\norm{\coefv_\si}_0$) explicitly in the cost function
results in a non-convex, discontinuous objective function. A common
procedure in sparse coding for this case is to estimate the optimum support
$\suppv_\si^\supsize$ for each possible support size,
$\supsize=1,2,\ldots,\natoms$.  Then, for each optimum support
$\{\suppv_\si^\supsize:\supsize\!=\!1,\ldots,\natoms\}$, \refeq{eq:total-codelength}
is solved in terms of the corresponding non-zero values of $\coefv_\si$,
yielding a candidate solution $\coefv_\si^\supsize$, and the one producing
the smallest codelength is assigned to $\opt{\coefv}_\si$. As an example of
this procedure, we propose Algorithm~\ref{alg:fss-mp}, which is a variant of
Matching Pursuit~\cite{mallat93}. As in \cite{mallat93}, we start with
$\coefv_\si^0=\vec{0}$, adding one new atom to the active set in each
iteration. However, instead of adding the atom that is maximally correlated
with the current residual $\errv$ to the active set, we add the one yielding
the largest decrease in overall codelength. The algorithm stops when no
further decrease in codelength is obtained by adding a new atom.  

An alternative for estimating
$\{\suppv_\si^\supsize:\supsize\!=\!1,\ldots,\natoms\}$ is to use a convex
model selection algorithm such as LARS/Lasso~\cite{efron04}, which also
begins with $\coefv_\si^0=\vec{0}$, adding one atom at a time to the
solution. For this case we propose to substitute the \cost{2} loss in
LARS/Lasso by $-\log \LG(\cdot)$, which is more consistent with
\refeq{eq:total-codelength} and can be approximated by the Huber loss
function~\cite{huber64}. This alternative will be discussed in detail in the
extended version of this work.

\begin{algorithm}[t]
\begin{scriptsize}
  \caption{\scriptsize Codelength-based Forward Selection.}
\label{alg:fss-mp}
\SetKw{Init}{initialize}
\SetKw{Set}{set}
\SetKw{Choose}{choose}
\SetCommentSty{textit}
\KwIn{Data sample $\datav$, dictionary $\dictm$}
\KwOut{The sparse code for $\datav$, $\coefv$}
\Init $\coefv \leftarrow \mat{0}; \errv \leftarrow \datav;  L \leftarrow L(\mat{0});\;\suppv \leftarrow \vec{0}$ \;
\Init $\corrv \leftarrow \dictm^T\errv$ \tcp*{correlation of current error with the dictionary}
\Repeat{$L_{\opt{\ai}} \geq L$}{
  \For{$\ai=1,\ldots,\natoms: \supp_\ai = 0$}{
    $\Delta_{\ai} \leftarrow [\corr_\ai]_{\delta_\coef}$  \tcp*{step $\Delta_{\ai}$ is correlation, quantized to prec. $\delta_\coef$}
    $L_{\ai} \leftarrow L(\coefv + \Delta_\ai \canon_\ai)$ \tcp*{$\canon_\ai$ is the $\ai$-th canonical vector of $\reals^{\natoms}$}

  }
  Choose $\opt{\ai} = \arg\min_{\ai: \supp_\ai = 0}\{L_\ai\}$\;
  \If{$L_{\opt{\ai}} < L$}{ 
    $L \leftarrow L_{\opt{\ai}}$ \tcp*{update current smallest codelength}
    $\supp_{\opt{\ai}} \leftarrow 1$
    \tcp*{update support vector}
    $\coefv \leftarrow \coefv + \Delta_{\opt{\ai}} \canon_{\opt{\ai}}$ 
    \tcp*{update coefficients vector}
    $\corrv \leftarrow \corrv - \Delta_{\opt{\ai}} \dictv_{\opt{\ai}}$
    \tcp*{update correlation}

  }
}
\end{scriptsize}
\end{algorithm}

\subsection{Dictionary learning}

Dictionary learning in the proposed framework proceeds in two stages. In the
first one, a maximum dictionary size $\natoms_{\mathrm{max}}$ is fixed and
the algorithm learns $\dictm$ using alternate minimization in $\coefm$ and
$\dictm$, as in standard dictionary learning approaches, now with the new
codelength-based metric. First, $\dictm$ is fixed and $\coefm$ is updated as
described in Section~\ref{sec:coding}. Second, keeping $\coefm$ fixed, the
update of $\dictm$ reduces to
$\arg\min_{\dictm}L_{\errrv}(\datam-\dictm\coefm)$ (recall that $L(\dictm)$
depends only on $\natoms$, thus being constant at this stage). We add the
constraint that $\norm{\dictv_\ai}_2 \leq 1, \ai=1,\ldots,\natoms$, a
standard form of dictionary regularization. Here too we approximate
$L_{\errrv}(\cdot)$ with the Huber function, obtaining a convex,
differentiable dictionary update step which can be efficiently solved using
scaled projected gradient. In practice, this function produces smaller
codelengths than using standard \cost{2}-based dictionary update, at the
same computational cost (see Section~\ref{sec:results}).

In the second stage, the size of the dictionary is optimized by pruning
atoms whose presence in $\dictm$ actually results in an increased average
codelength. In practice, the final size of the dictionary reflects the
intuitive complexity of the data to be modeled, thus validating the approach
(see examples in Section~\ref{sec:results}).

\section{Results and conclusion}
\label{sec:results}

The first experiment assesses that the learned models produce compressed
descriptions of natural images. For this, we adapted a dictionary to the
Pascal'06 image
database\footnote{\url{http://pascallin.ecs.soton.ac.uk/challenges/VOC/databases.html}},
and encoded several of its images. The average bits per pixel obtained was
$4.08$ bits per pixel (bpp), with $\natoms=250$ atoms in the final
dictionary. We repeated this using \cost{2} instead of Huber loss, obtaining
$4.12$ bpp and $\natoms=245$.

We now show example results obtained with our framework in two very
different applications. In both cases we exploit spatial correlation between
codes of adjacent patches by learning Markovian dependencies between their
supports (see \cite{carin10} for related work in the Bayesian framework).
More specifically, we condition the probability of occurrence of an atom at
a given position in the image, on the occurrence of that same atom in the
left, top, and top-left patches. Note that, in both applications, the
results were obtained without the need to adjust any parameter (although
$\delta_\coef$ is a parameter of Algorithm~\ref{alg:fss-mp}, it was fixed
beforehand to a value that did not introduce noticeable additional
distortion in the model, and left untouched for the experiments shown here).

The first task is to estimate a clean image from an observed noisy version
corrupted by Gaussian noise of known variance $\sigma^2_\noiserv$. Here
$\datam$ contains all (overlapping) $8{\times}8$ patches from the noisy
image. First, a dictionary $\dictm$ is learned from the noisy patches. Then
each patch is encoded using $\dictm$ with a denoising variant of
Algorithm~\ref{alg:fss-mp}, where the stopping criterion is changed for the
requirement that the observed distortion falls within a ball of radius
$\sqrt{\ndims}\sigma_\noiserv$. Finally, the estimated patches are overlapped again and
averaged to form the denoised image. From the results shown in
Figure~\ref{fig:denoising}, it can be observed that adding a Markovian
dependency between patches consistently improves the results.  Note also
that in contrast with \cite{aharon06}, these results are fully parameter
free, while those in \cite{aharon06} significantly depend on carefully tuned
parameters.

\begin{figure}
\centering{
\includegraphics[height=0.8in]{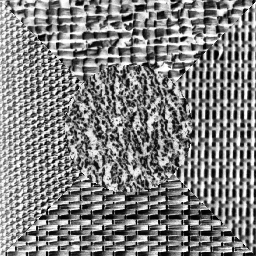}\hspace{1.5pt}%
\includegraphics[height=0.8in]{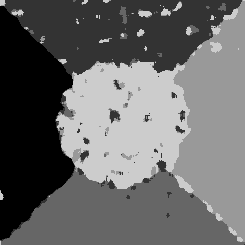}\hspace{1.5pt}%
\includegraphics[height=0.8in]{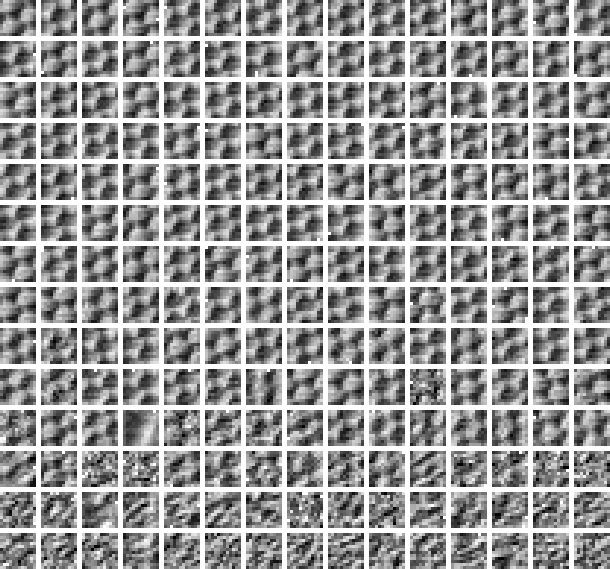}\hspace{1.5pt}%
\includegraphics[height=0.8in]{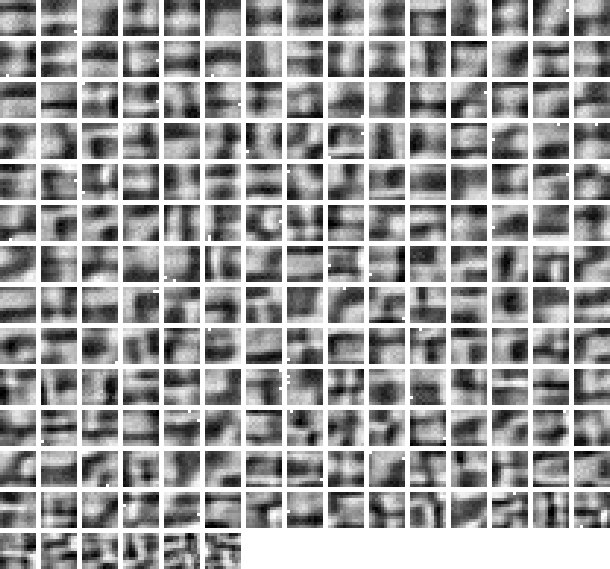}%
}\caption{\label{fig:texture}Segmentation
  results. Left to right: sample ``Nat-5c'' taken from
  {\scriptsize\url{http://www.ux.uis.no/~tranden/data.html}}, obtained
  segmentation (error rate $7.5\%$), learned dictionaries for classes $1$ ($\natoms_1=210$) and $2$ ($\natoms_2=201$).}
\end{figure}

The second application is texture segmentation via patch
classification. Here we are given $\nclasses$ images with sample textures,
and a target mosaic of textures, and the task is to assign each pixel in the
mosaic to one of the textures. Again, all images are decomposed into
overlapping patches. This time a dictionary $\dictm^r$ is learned for each
texture $r=1,\ldots,\nclasses$ using the patches from the training
images. Then, each patch in the mosaic is encoded using all available
dictionaries, and its center pixel is assigned to the class which produced
the shortest description length for that patch. The final result includes a
simple $3{\times}3$ median filter to smooth the segmentation. We show a
sample result in Figure~\ref{fig:texture}. Here, again, the whole process is
parameter free, and adding Markovian dependency improves the overall error
rate ($7.5\%$ against $8.5\%$).

In summary, we have presented an MDL-based sparse modeling framework, which
automatically adapts to the inherent complexity of the data at hand using
codelength as a metric. As a result, the framework can be applied
out-of-the-box to very different applications, obtaining competitive results
in all the cases presented. We have also shown how prior information, such
as spatial dependencies, are easily added to the framework by means of
probability models.

\renewcommand{\best}[1]{#1} 
\begin{figure}
\begin{center}
\begin{minipage}[t]{1.85in}\centering%
\setlength\tabcolsep{1pt}
\resizebox{1.85in}{0.75in}{%
\begin{tabular}[b]{|l|ccc|ccc|}\hline
noise $\rightarrow$ & \multicolumn{3}{|c}{$\sigma_\errrv=10$}& 
      \multicolumn{3}{|c|}{$\sigma_\errrv=20$}\\
image $\downarrow$ & IID & Markov & \cite{aharon06}&
      IID & Markov & \cite{aharon06} \\\hline
lena &     35.0 & 35.0 & \best{35.5} & 32.0 & 32.3 & \best{32.4} \\
barbara &  33.9 & 34.1 & \best{34.4} & 30.6 & 30.7 & \best{30.8} \\
boats &    32.9 & 32.9 & \best{33.6} & 30.1 & 30.2 & \best{30.3} \\
peppers &  34.0 & 34.0 & \best{34.3} & 31.4 & \best{31.5} & 30.8 \\\hline
\end{tabular}%
}%
\end{minipage}\hspace{1pt}%
\begin{minipage}[t]{0.75in}\centering%
\includegraphics[height=0.75in]{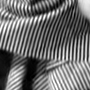}%
\end{minipage}\hspace{1pt}%
\begin{minipage}[t]{0.75in}\centering%
\includegraphics[height=0.75in]{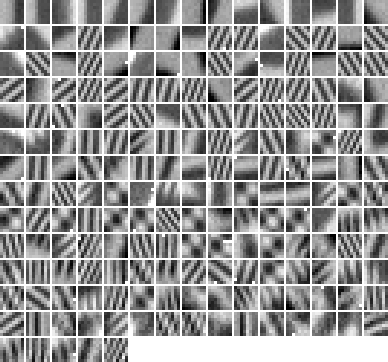}%
\end{minipage}
\caption{\label{fig:denoising}Denoising results. Left to right: PSNR of denoised images for different methods and noise levels (including \cite{aharon06} as a reference), sample recovered patch from Barbara, best dictionary for Barbara ($\natoms=200$, we used an initial $\natoms_{\max}=512$). }
\end{center}
\end{figure}

\bibliographystyle{IEEEbib}
\bibliography{icassp2011}

\end{document}